\documentclass[12pt]{article}
\usepackage{epsfig}
\usepackage{amsmath}
\usepackage{hhline}
\usepackage{amssymb}
\usepackage{times}
\usepackage{cite}
\usepackage{rotating}

\newlength{\dinwidth}
\newlength{\dinmargin}
\newcommand{\siggp}{$\sigma_{\gamma p}$}
\newcommand{\alprim}{\alpha^\prime}
\newcommand{\gpge}{$ep\rightarrow e\gamma p$}
\newcommand{\wwa}{Weizs\"acker-Williams approximation}

\newcommand{\ups}{$\Upsilon$}
\newcommand{\upsts}{$\Upsilon(2S$)}
\newcommand{\psits}{$\psi(2S$)}
\newcommand{\upsths}{$\Upsilon(3S$)}
\newcommand{\upsos}{$\Upsilon(1S$)}
\newcommand{\cms}{centre-of-mass}
\newcommand{\jpsi}{$J/\psi$}

\newcommand{\cs}{cross section}
\newcommand{\dsdt}{$d\sigma/dt$}
\newcommand{\ggmm}{$\gamma \gamma \rightarrow \mu^+\mu^-$}
\newcommand{\gpcs}{$\gamma p$ cross section}
\newcommand{\wgp}{$W_{\gamma p}$}
\newcommand{\wgpw}{W_{\gamma p}}

\newcommand{\GeV}{\,\mbox{GeV}}

\newcommand{\ra}{\rightarrow}

\newcommand{\gevt}{\,\mbox{\GeV$^2$}}
\newcommand{\gev}{\,\mbox{GeV}}
\newcommand{\gmt}{\,\mbox{\GeV$^{-2}$}}
\newcommand{\ccbar}{c\overline{c}}

\def\lsim{\mathrel{\rlap{\lower4pt\hbox{\hskip1pt$\sim$}}
    \raise1pt\hbox{$<$}}}                % less than or approx. symbol
\def\gsim{\mathrel{\rlap{\lower4pt\hbox{\hskip1pt$\sim$}}
    \raise1pt\hbox{$>$}}}                % greater than or approx. symbol
\newcommand{\qsq}{\ensuremath{Q^2} }
\newcommand{\gevsq}{\ensuremath{\mathrm{GeV}^2} }
\newcommand{\gevmsq}{\ensuremath{\mathrm{GeV}^{-2}} }

\newcommand{\gp}{\ensuremath{\gamma}p }

\begin{document}
%\psdraft
% Journal macro
\def\Journal#1#2#3#4{{#1} {\bf #2} (#3) #4}
\def\NCA{\em Nuovo Cimento}
\def\NIM{\em Nucl. Instrum. Methods}
\def\NIMA{{\em Nucl. Instrum. Methods} {\bf A}}
\def\NPB{{\em Nucl. Phys.}   {\bf B}}
\def\PLB{{\em Phys. Lett.}   {\bf B}}
\def\PRL{\em Phys. Rev. Lett.}
\def\PRD{{\em Phys. Rev.}    {\bf D}}
\def\ZPC{{\em Z. Phys.}      {\bf C}}
\def\EJC{{\em Eur. Phys. J.} {\bf C}}
\def\CPC{\em Comp. Phys. Commun.}
\def\ib {\em ibid.}

\begin{titlepage}
%\noindent
%Date: \today \\
%Version:  \the\time     \\
%Version:  3.0    \\
%Editors:  P. Merkel, B. Naroska \\
%Referees: D. Reyna, T. Sloan \\
%Comments by February 18 th, 2000
\begin{flushleft}
  {\tt DESY 00-037} \hfill {\tt ISSN 0418-9833} \\
  {\tt March 2000}  
\end{flushleft}

\vspace*{3cm}

\begin{center}

\begin{Large}

\noindent

%\begin{center}
%\begin{LARGE} {DRAFT}\end{LARGE}
%\begin{Large}
\vspace{2cm}

{\bf Elastic Photoproduction of \boldmath$J/\psi$ and $\Upsilon$ Mesons at HERA}

\vspace{2cm}

\vspace{1cm}

H1 Collaboration
\end{Large}

\vspace{2cm}

%%%%%%%%%%%%%%%%%%%%%%%%%%%%%%%%%%%%%%%%%%%%%%%%%%%%%%%%%%%%%%%%%%%%%%%%%%%%%%%%%%%%%%%%%%%%%%%%%%%

\begin{abstract}

Cross sections for elastic  photoproduction of \jpsi\ and \ups\ mesons are presented.
For \jpsi\ mesons the dependence on the photon-proton centre-of-mass energy \wgp\ is 
analysed in an extended range with respect to previous measurements of 
$26\leq\wgpw \leq 285\,\gev$. The  measured energy dependence is parameterized as
\siggp$\propto\wgpw^\delta$ with $\delta=0.83\pm0.07$. The differential \cs\ 
$d\sigma/dt$ for \jpsi\ mesons is derived, its dependence on \wgp\ and on $t$ is 
analysed and the effective trajectory (in terms of Regge theory) is determined to be 
$ \alpha(t)= (1.27\pm0.05)+(0.08\pm0.17)\cdot t/\gevsq$. Models based on perturbative 
QCD and on pomeron exchange are compared to the data.

\end{abstract}

\vspace{1.5cm}

\begin{center}
{\sl Submitted to Phys.~Lett.~B}
\end{center}

\end{center}
\end{titlepage}

%%%%%%%%%%%%%%%%%%%%%%%%%%%%%%%%%%%%%%%%%%%%%%%%%%%%%%%%%%%%%%%%%%%%%%%%%%%%%%%%%%%%%%%%%%%%%%%%%%%

%   H1AUTS  Author list by names, no. of authors  341
%           status: 11/01/:0   11.05.25
 C.~Adloff$^{33}$,                %WUPP-ST                  Adloff             
 V.~Andreev$^{24}$,               %LPI -PD                  Andreev            
 B.~Andrieu$^{27}$,               %ECPL-PD                  Andrieu            
 V.~Arkadov$^{35}$,               %ZEUT-ST    10/96         Arkadov            
 A.~Astvatsatourov$^{35}$,        %ZEUT-ST     02/98        Astvatsatourov     
 I.~Ayyaz$^{28}$,                 %PARI-LEFT  12/99         Ayyaz              
 A.~Babaev$^{23}$,                %ITEP-PD                  Babaev             
 J.~B\"ahr$^{35}$,                %ZEUT-PD                  Baehr              
 P.~Baranov$^{24}$,               %LPI -PD                  Baranovp           
 E.~Barrelet$^{28}$
% ...

\vspace{1cm}

%     H1 Institutes as appearing on publications
% ...
$ ^{23}$ Institute for Theoretical and Experimental Physics,
          Moscow, Russia \\
 $ ^{24}$ Lebedev Physical Institute, Moscow, Russia$^{e,h}$ \\
% ...
$ ^{27}$ LPNHE, \'{E}cole Polytechnique, IN2P3-CNRS, Palaiseau, France \\
 $ ^{28}$ LPNHE, Universit\'{e}s Paris VI and VII, IN2P3-CNRS,
          Paris, France \\
% ...
$ ^{33}$ Fachbereich Physik, Bergische Universit\"at Gesamthochschule
          Wuppertal, Wuppertal, Germany$^a$ \\
% ...
$ ^{35}$ DESY, Zeuthen, Germany$^a$ \\
% ...
$ ^a$ Supported by the Bundesministerium f\"ur Bildung, Wissenschaft,
        Forschung und Technologie, FRG,
        under contract numbers 7AC17P, 7AC47P, 7DO55P, 7HH17I, 7HH27P,
        7HD17P, 7HD27P, 7KI17I, 6MP17I and 7WT87P \\
% ...
$ ^e$ Supported by the Deutsche Forschungsgemeinschaft \\
% ...
$ ^h$ Supported by Russian Foundation for Basic Research 
       grant no. 96-02-00019 \\
% ...
%Supported by the Alexander von Humboldt Foundation \\
% $ ^{m}$ Foundation for Polish Science fellow \\

%%%%%%%%%%%%%%%%%%%%%%%%%%%%%%%%%%%%%%%%%%%%%%%%%%%%%%%%%%%%%%%%%%%%%%%%%%%%%%%%%%%%%%%%%%%%%%%%%%%
\newpage
\section{Introduction} 

A large contribution to the photoproduction of \jpsi\ mesons is the {\em elastic} 
process, $\gamma\,p\ra J/\psi\,p$, which has been studied previously over a wide 
range, from threshold up to a photon-proton \cms\ energy, \wgp, of approximately 
$160\,\gev$ in $ep$ collisions at HERA.  The cross 
section for elastic \jpsi\ photoproduction was observed to rise much more
steeply with \wgp~\cite{H1962, Zeus971, H1991, Zeus991} than that for the light
vector mesons~\cite{oldrho}. 
In perturbative quantum chromodynamics (pQCD), 
the process is viewed as an almost real photon from the 
lepton coupling to a pair of $\ccbar$\ quarks which interact with the 
proton via exchange of two (or more) gluons and evolve into a \jpsi\
meson.  
The cross section is then proportional to the square of 
the gluon density of the proton. The rapid rise of the gluon density at
low fractional gluon momenta leads to the observed rise of the \cs\ with
$W_{\gamma p}$~\cite{Ryskin,F_K_S,F_K_S97}. 
In such calculations, the heavy 
quark mass serves as the QCD scale. Therefore, it is of interest to measure the  
\cs\ of the $\Upsilon$ meson, for which perturbation theory might be more reliable. 
In non-perturbative models, $J/\psi$ production has been described in terms of 
pomeron exchange~\cite{collins,1pom}. Since the standard soft pomeron fails to 
describe inclusive electroproduction and the observed \wgp\ dependence of the \jpsi\ 
\cs, extensions of such models have been proposed~\cite{refalt}. In one such 
extension~\cite{2pom}, which is compared to the data of this paper, an extra 
``hard'' pomeron is introduced. 

Here, we present new data obtained by the H1 experiment which may lead to a better 
understanding of the elastic production process. The \wgp\ dependence of the \cs\ for 
elastic \jpsi\ photoproduction is measured, with increased statistics in an enlarged 
range of $26\leq\wgpw\leq285\,\gev$, and compared to QCD and pomeron models. 
Furthermore, the differential cross section \dsdt, where $|t|$ is the squared 
four-momentum transfer to the proton, is measured as a function of $t$ and of \wgp. 
From its \wgp\ dependence, an effective Regge trajectory is determined using data 
from this experiment alone, thus avoiding relative normalization problems when 
comparing data from different experiments. The results are compared to the predictions 
of the two-pomeron model~\cite{2pom} and also to a recent calculation in 
next-to-leading order (NLO) using the BFKL equation~\cite{bfklpom}. Finally, we 
present a measurement of elastic photoproduction of \ups\ mesons which allows a test 
of the predictions of pQCD at a higher mass scale~\cite{F_S_M99,MRT99}.

%%%%%%%%%%%%%%%%%%%%%%%%%%%%%%%%%%%%%%%%%%%%%%%%%%%%%%%%%%%%%%%%%%%%%%%%%%%%%%%%%%%%%%%%%%%%%%%%%%%

\section{Data Analysis}

We report, here, an analysis of the process $ep\rightarrow eVp$, ($V=J/\psi \ \mbox{or} 
\ \Upsilon$), where \jpsi\ mesons decaying to $\mu^+\mu^-$ and $e^+e^-$ and 
\ups$\ra\mu^+\mu^-$ are observed. The data were taken with the H1 detector while 
HERA was operating with positrons of $27.5\,\gev$ and protons of $820\,\gev$. 

The kinematics of an $ep$ reaction are described by the square of the $ep$ 
centre-of-mass energy $s = (p+k)^2$, the squared four-momentum transfer from the 
positron $Q^2 = -q^2= -(k-k^\prime)^2$ and the scaled energy transfer $y=(p\cdot q) / 
(p \cdot k) $, where $k$,  $k^\prime$, $p$ and $q$ are the four-momenta of the 
incident and scattered positron, the incoming proton and the exchanged photon, 
respectively. In the photoproduction domain \qsq$\rightarrow 0$, the positron is 
scattered at small angles and generally not observed. In the present analysis, the 
photoproduction region is defined by \qsq$\leq 1\,\gevsq$. The average \qsq\ values, 
derived from the Monte Carlo simulation, are $\langle\qsq \rangle\simeq 0.05\,\gevsq$ 
for the \jpsi\ analyses and $\langle\qsq\rangle\simeq 0.11\,\gevsq$ for the \ups\ 
analysis. In the photoproduction limit, the square of the \cms\ energy of the 
photon-proton system $\wgpw^2 = (p+q)^2$ is  given by $\wgpw^2 \simeq ys$. It is 
calculated from the relation: $y = (E-p_z)_{V}/(2\,E_e)$, where $E$ and $p_z$ denote 
the total energy and longitudinal component\footnote{The forward ($+z$) direction, with 
respect to which polar angles are measured, is defined as that of the incident 
proton beam direction.} of the momentum of the vector meson $V$ 
and $E_e$ is the energy of the incident positron. 

%%%%%%%%%%%%%%%%%%%%%%%%%%%%%%%%%%%%%%%%%%%%%%%%%%%%%%%%%%%%%%%%%%%%%%%%%%%%%%%%%%%%%%%%%%%%%%%%%%%

\subsection*{Detector and Data Selection }
\label{detector}

The H1 detector is described in detail elsewhere~\cite{h1det}. The main detector 
consists of a system of drift and proportional chambers: the central track detector 
(CTD), covering the polar angle range $15^\circ \lsim \theta \lsim 165^\circ$ and 
the forward track 
detectors (FTD), covering $7^\circ \lsim\theta \lsim 25^\circ$. The tracking detectors 
are surrounded by a liquid argon calorimeter (LAr, $4^\circ \lsim \theta \lsim 
154^\circ$), a scintillating fibre calorimeter (spaghetti calorimeter, 
SpaCal~\cite{spacal}, $153^\circ \lsim \theta \lsim 178^\circ$) and an instrumented 
iron return yoke (central muon detector, CMD, $4^\circ \lsim \theta \lsim 171^\circ$). 
The interaction region is surrounded by an assembly of high precision silicon 
detectors, the backward part of which (backward silicon tracker, BST~\cite{bst}) is 
used in this analysis. The BST, which became fully operational in 1997, is a four 
plane silicon strip detector telescope arranged concentrically around the beam axis 
to measure the polar angle of tracks ($172^\circ \lsim \theta \lsim 176.5^\circ$).

Electrons are identified in the LAr or the SpaCal calorimeters. Muons are identified 
in the LAr calorimeter, in the CMD and, at small polar angles, in the forward muon 
detector (FMD, $3^\circ \lsim \theta \lsim 17^\circ$), which consists of drift 
chambers on either side of a toroidal magnet. The momentum and angular measurements 
of the decay leptons are provided by the CTD (FTD) in the central (forward) detector 
region. At the highest values of \wgp, both electrons are scattered into SpaCal. In 
this case, the polar angle and the event vertex are derived from track measurements 
in the BST. 

In elastic photoproduction of \jpsi\ and \ups\ mesons, both the scattered positron 
and the scattered proton are not generally detected. The largest background is due 
to processes in which the proton dissociates to form a hadronic state $Y$, the proton 
remnant. To reject such events, the so-called ``forward detectors'' are utilized:  
the proton remnant tagger, PRT, scintillation counters covering polar angles close to 
the proton beam, the drift chambers of the FMD in front of the toroid and the forward 
part of the LAr calorimeter ($\theta \leq 10^\circ$). Proton remnants with masses 
$M_Y\gsim 1.6\mbox{~GeV}$ can be detected.

The experimental signature of the process $e\,p\rightarrow e\,V\,p$, $V\ra 
\ell^+\ell^-$ in the photoproduction domain is a pair of leptons in an otherwise 
``empty'' detector. The data selection thus requires only two decay leptons to be 
observed in the main detector with no signal in the forward detectors. The value of 
\wgp\ is related to the polar angles of the vector meson and therefore also to the 
polar angle of the decay leptons. Selection criteria are used which 
correspond to different detector regions and ranges of \wgp\ (see Table~\ref{tab1}). 
For each analysis region, the decay mode, $\mu^+\mu^-$ or $e^+e^-$, with the better 
detection efficiency is chosen.      

%%%%%%%%%%%%%%%%%%%%%%%%%%%%%%%%%%%%%%%%%%%%%%%%%%%%%%%%%%%%%%%%%%%%%%%%%%%%%%%%%%%%%%%%%%%%%%%%%%%

\newcommand{\rb}[1]{\raisebox{1.5ex}[-1.5ex]{#1}}
\renewcommand{\arraystretch}{1.1} 
\begin{table}[ht]
\begin{tabular}{|l|c|c|c|c|c|}\hline

                                       &
\multicolumn{2}{c|}{Central region}    & 
\multicolumn{2}{c|}{Backward region}   & 
Forward region                         \\ \hline\hline 

Decay channel                                      &
$J/\psi\rightarrow\mu^+\mu^-$                      & 
$\Upsilon\rightarrow\mu^+\mu^-$                    & 
\multicolumn{2}{c|}{$J/\psi\rightarrow e^+e^-$}    & 
$J/\psi\rightarrow\mu^+\mu^-$                      \\ \hline

$\langle Q^2\rangle$ \ \ \ \ \ \: $[$GeV$^2]$ &
$0.05$                                           & 
$0.11$                                           & 
\multicolumn{2}{c|}{$0.05$}                      & 
$0.05$                                           \\ \hline

Lepton polar                           &
                                       & 
                                       & 
$80 - 155$&$155 - 175.5$     &$8 - 16$                         \\

angle region \ \ $[^\circ]$            &
\raisebox{1.5ex}[0cm][0cm]{$20 - 160$}   & 
\raisebox{1.5ex}[0cm][0cm]{$20 - 165$} &$155 - 176$& 
                {$155 - 174$} & $20 - 160$                          \\ \hline

$W_{\gamma p}\,$range$\,[$GeV$]$ &
$40 - 150$                       & 
$70 - 250$                       & 
$135 - 210$                      &
$210-285$                        & 
$26 - 36$                        \\ \hline

Data from    &
$1996 - 97$  &
$1994 - 97$  &
$1996 - 97$  &
$1997$         &
$1996 - 97$  \\ \hline

$\int\hspace{-0.1cm}{\cal L}\:\mbox{dt}$ \ \ \ \ \ \: $[$pb$^{-1}]$ &
$20.5$                                                                 &
$27.5$                                                                 &
$20.5$                                                                 &
$10.0$                                                                 &
$20.5$                                                                 \\ \hline
\end{tabular}
 \caption{\em Summary of the different data sets.} 
\label{tab1}
\end{table}

%%%%%%%%%%%%%%%%%%%%%%%%%%%%%%%%%%%%%%%%%%%%%%%%%%%%%%%%%%%%%%%%%%%%%%%%%%%%%%%%%%%%%%%%%%%%%%%%%%%

In the central region, $J/\psi\rightarrow\mu^+\mu^-$ is analysed in the range 
$40\leq \wgpw\leq 150\,\gev$. Exactly two oppositely charged particles, measured in 
the CTD, are required with transverse momenta (with respect to the beamline) $p_t > 
0.8\;\mbox{GeV}$. At least one of these must be validated as a muon in the LAr 
calorimeter or in the CMD (for further details see \cite{H1962,H1991,merkel}). 
Background from cosmic ray muons is rejected by an acolinearity cut. Triggers based 
on muon and track signatures from the decay leptons are used. For the reconstruction 
of \ups\ decays, the acceptance region is $70\leq\wgpw\leq 250\,\gev$. In order to 
optimize the signal to background ratio, both muons must be identified and a 
tighter cut against cosmic rays is applied.

For the $J/\psi$ analysis in the backward region, the decay to electrons is used. Two 
sets of selection criteria are applied, depending on the event kinematics. The first 
set ($135<W_{\gamma p}<210\:\mbox{GeV}$) requires one decay electron to be measured 
in the CTD and one in the SpaCal calorimeter. The electron measured in the CTD must 
have a momentum $p>0.8\,\gev$ and a transverse momentum $p_t>0.7\,\gev$. It is 
validated by an electromagnetic cluster in the LAr calorimeter. The other electron is 
selected by requiring a cluster in the SpaCal calorimeter with energy above 
$4.2\,\gev$, which is assumed to originate from the intersection of the first 
electron track with the $z$-axis. In order to suppress inelastic contributions and 
background, no further tracks are allowed and any energy in the SpaCal outside the 
selected electron cluster must be below $6\,\%$ of its energy. The second selection 
($210<W_{\gamma p}<285\:\mbox{GeV}$) requires both electrons to be detected in the 
SpaCal calorimeter as described above. They must be validated either by hits in a 
proportional chamber or by a track in the BST. By requiring two charged particles, at 
least one of which is in the acceptance of the BST, most of the non-resonant 
background from the QED process \gpge\ is rejected. The triggers for both selections 
are based on signals from the SpaCal and the CTD. In addition the triggers in the 
central and backward regions use a trigger based on neural networks~\cite{L2NN}.  

In the $J/\psi$ analysis in the forward region ($26 \leq \wgpw \leq 36\mbox{~GeV}$), 
the FMD is used to identify one decay muon with momentum above $5\,\gev$. The other 
muon is measured in the CTD with momentum above $0.8\,\gev$. No other tracks, except 
those associated with the muons, are allowed. Proton dissociative events are rejected 
by a forward detector analysis, taking into account that one of the muons passes
through the drift chambers of the FMD and may pass through the forward part of the 
LAr calorimeter. The FMD also supplies the trigger for these events in conjunction 
with signals from track chambers and the central muon detector. 

\begin{figure}[tb] \centering
  \setlength{\unitlength}{1cm}
  \begin{picture}(17.,5.0)
    \put(-2.3,-2.3){\epsfig{file=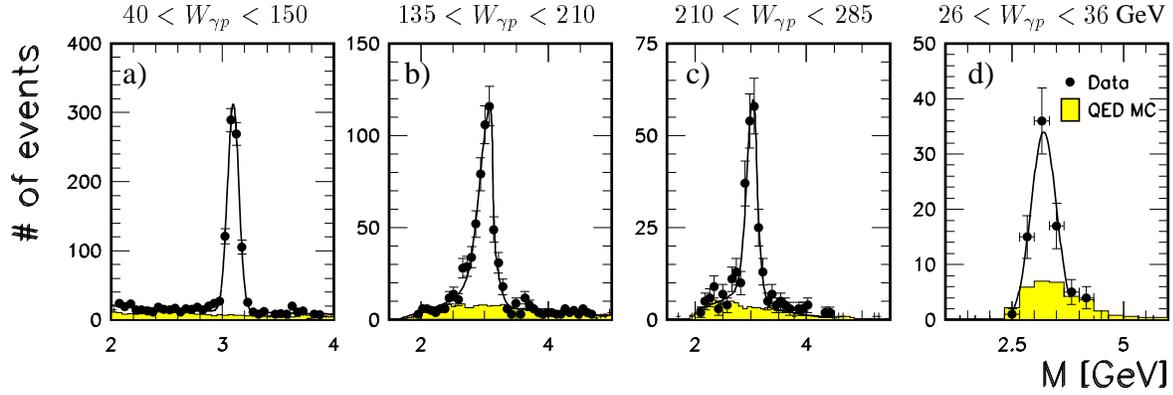,width=21cm}}
  \end{picture} 
  \caption{\em Mass spectra for the four \wgp\ regions of the elastic $J/\psi$
           selection: a) $\mu^+\mu^-$ pairs in the central region, b) and c) 
           $e^+e^-$ pairs in the backward region, d) $\mu^+\mu^-$ pairs in the 
           forward region. A fit of the signal region as described in the text 
           (full line) and the simulated mass spectra from QED background processes 
           (shaded histograms) are shown.} 
 \label{fig1}
\end{figure} 

%%%%%%%%%%%%%%%%%%%%%%%%%%%%%%%%%%%%%%%%%%%%%%%%%%%%%%%%%%%%%%%%%%%%%%%%%%%%%%%%%%%%%%%%%%%%%%%%%%%

\subsection*{Cross Section Determination}
\label{crosssection}

After the data selection, clear peaks at the $J/\psi$ mass are observed in the 
distributions of the invariant mass of the two leptons in all analysis regions 
(Figure~\ref{fig1}). The remaining background in the forward and central regions 
is dominated by the process \ggmm, where the photons originate from both the positron 
and proton. In the backward region at very high \wgp, the background is composed of 
the processes $\gamma\gamma\rightarrow e^+e^-$ and \gpge, where the photon and 
electron are scattered at large angles. These background processes can be simulated 
by the LPAIR~\cite{lpair} and COMPTON~\cite{compton} generators, respectively.

The number of $J/\psi$ events in the central region (Figure~\ref{fig1}a) is 
determined in each analysis bin by a fit which uses a Gaussian for the signal and 
a polynomial for the background. In the backward region (\jpsi$\ra e^+e^-$), the 
number of events is determined by fitting the signal region with a Gaussian and an 
exponential to describe the low mass tail, and subtracting the simulated background 
(Figure~\ref{fig1}b,c). The low mass tail is caused by radiative effects in the
the detector  material. In the forward region, the background is determined by the 
Monte Carlo simulation.

For the calculation of the \cs, the number of events without a signal in the forward 
detectors is corrected for acceptance and efficiency losses by using the Monte Carlo 
simulation DIFFVM~\cite{diffvm}. DIFFVM is based on the Vector Meson Dominance Model 
and Regge phenomenology. Either the elastic or the proton dissociative processes 
can be generated. The $Q^2$, \wgp\ and $t$ dependences have been adjusted to fit the 
data.

A correction is applied for the remaining background from proton dissociation 
which is typically $10-13\%$ in the 
central and backward regions and $30\%$ in the forward region. It is determined from 
the Monte Carlo simulation and cross checked with data samples with and without 
signals in the forward detectors. The non-prompt $J/\psi$ events from \psits\ decays 
($\sim3\%$) are subtracted according to their measured cross section~\cite{psi2s} and 
branching ratio~\cite{PDG98}. The total efficiencies (including acceptance) for the 
\jpsi\ analyses are typically $5\%$, $14\%$ and $20\%$ in the forward, central and 
backward regions, respectively. The trigger efficiencies are about $18\%$, $20\%$ 
and $38-60\%$, depending on \wgp. At the highest \wgp\ values 
($W_{\gamma p}\gsim 200\:\mbox{GeV}$), the trigger efficiency is $100\%$, within 
errors. The losses in this region are mainly due to the limited acceptance.
 
The $ep$ cross section is obtained using the measured luminosity and the branching 
ratio of the \jpsi\ decay to electrons or muons, $(6.0\pm0.2)\%$~\cite{PDG98} each. 
For the decay to electrons, the radiative decay $J/\psi\rightarrow e^+e^-\gamma$ 
with the branching ratio $(0.88\pm 0.14)\%$~\cite{PDG98} is also taken into account. The 
photoproduction cross section is calculated from the electroproduction cross section, 
assuming factorization of the $ep$ reaction into emission of photons, described by a 
photon flux (\wwa~\cite{WWA}), followed by a $\gamma p$ reaction.

The systematic error for the \jpsi\ analysis in the central region increases with 
\wgp\ from $11\%$ to $14\%$ with an estimated correlated part, which affects only 
the normalization, of $5-6\%$. The largest contributions are the uncertainties in the 
determination of the proton dissociative background ($5-10\%$), the efficiency of the 
lepton identification ($6.5\%$) and the trigger efficiency ($6\%$). In the backward 
region, the systematic uncertainties are estimated to be, on average, $16\%$. For 
$W_{\gamma p}\lsim 200\:\mbox{GeV}$, the contributions are similar to those in 
the central region. For $W_{\gamma p}\gsim 200\:\mbox{GeV}$, the main contributions 
are due to the uncertainties in the subtraction of the residual background from QED 
processes ($10\%$) and proton dissociation ($10\%$). In the forward region (low \wgp), 
the total uncertainty is $26\%$, where the largest contributions are the uncertainty 
of the trigger efficiency ($18\%$) and the determination of the background ($16\%$).

Radiative corrections have been neglected throughout the analysis. Recent calculations 
for elastic photoproduction of $\rho$ mesons~\cite{aku} suggest that, for 
\jpsi\ photoproduction, they are small compared to experimental errors.

%%%%%%%%%%%%%%%%%%%%%%%%%%%%%%%%%%%%%%%%%%%%%%%%%%%%%%%%%%%%%%%%%%%%%%%%%%%%%%%%%%%%%%%%%%%%%%%%%%%

\section{\boldmath{\wgp}\ Dependence of \boldmath{$J/\psi$} Production}

\begin{figure}[htb] \centering
  \setlength{\unitlength}{1cm}
  \begin{picture}(16.,9.)
    \put(-1.4,0.8){\epsfig{file=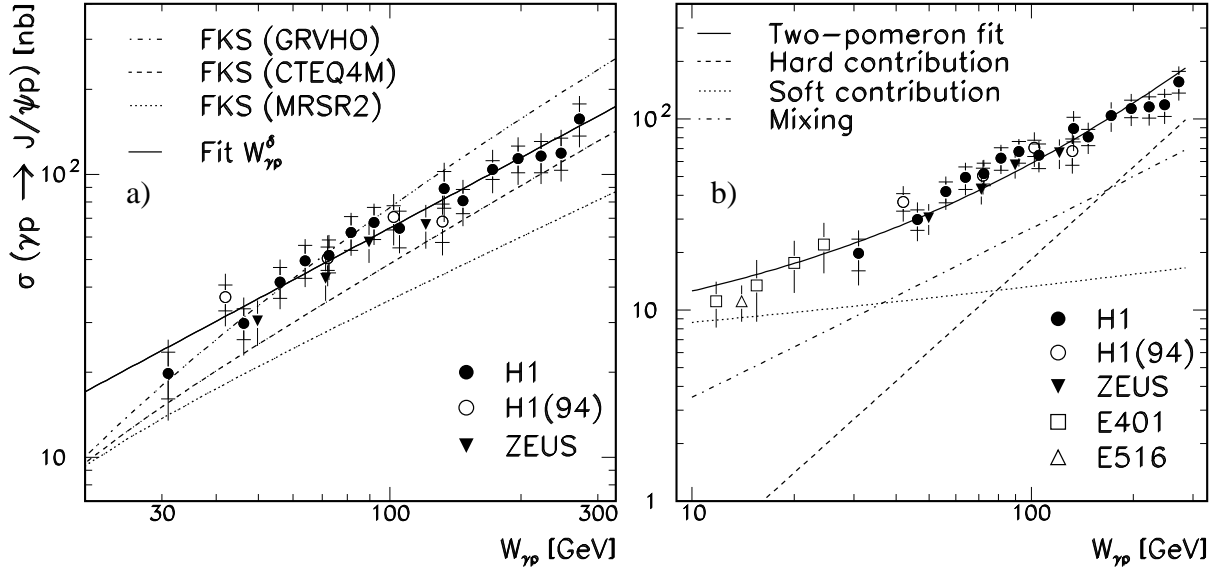,width=18.cm}}
  \end{picture}
\vspace{-2.3cm}  
  \caption{\em a) The cross section $\sigma (\gamma p \rightarrow J/\psi p)$ 
           versus $W_{\gamma p}$ from this analysis and previous HERA results. The 
           inner and outer error bars show the statistical and total errors, 
           respectively. The full line represents a fit of the form 
           \protect\siggp$\propto \wgpw^\delta$ to the present H1 data, yielding 
           $\delta=0.83\pm0.07$ with a $\chi^2/\mathit{dof}=4.6/13$. QCD predictions 
           by Frankfurt et al.~\protect\cite{F_K_S97} using various  
           parameterizations of the gluon density in the proton and the charm quark 
           mass of $1.48\:\mbox{GeV}$ as chosen in \cite{F_K_S97} are also shown.
           b) The same HERA data as in a) and results from fixed-target experiments 
           with a fit of the two-pomeron model by Donnachie and Landshoff 
           (full line; $\chi^2/\mathit{dof}=16.1/22$). Only the contributions of the 
           soft and hard pomeron amplitudes were adjusted allowing for 
           mixing~\cite{2pom}. The separate contributions are indicated.}
  \label{fig2}
\end{figure}

The cross section for elastic photoproduction of $J/\psi$ mesons, as a function of 
$W_{\gamma p}$, is listed in Table~\ref{tab2} and is shown in 
Figure~\ref{fig2}a, together with previous H1 and ZEUS results~\cite{H1962,Zeus971}. 
The data show good agreement in the region of overlap. A fit of the form 
$W_{\gp}^{\delta}$ to the present H1 data between $26$ and $285\:\mbox{GeV}$ yields a 
value of $\delta=0.83\pm 0.07$ (including statistical and systematic errors). This 
result confirms, with smaller errors, the steep rise of the \jpsi\ \cs\ with \wgp\ 
observed previously~\cite{H1962,Zeus971}, which is much larger than for the light 
vector mesons~\cite{oldrho}.

In Figure~\ref{fig2}a, the HERA data are shown with predictions from a leading order 
pQCD calculation~\cite{F_K_S,F_K_S97} using various gluon density distributions. 
The important prediction concerns the \wgp\ dependence of the cross section since the 
absolute magnitude is more dependent on a number of parameters and on non-perturbative 
effects. The slope of the data is described well using either the gluon density 
CTEQ4M~\cite{CTEQ4} or MRSR2~\cite{MRSR} while the GRV-HO~\cite{GRV} based result 
is too steep.

For the comparison with models based on pomeron exchange, the present data, previous 
HERA data and results from  fixed-target experiments~\cite{FTPSpsi}\footnote{Only 
fixed-target results are shown which were performed on $H_2$ or $D_2$ targets and 
which have been corrected for contributions from proton dissociative processes.} are 
used (Figure~\ref{fig2}b). A pomeron trajectory of the form $\alpha(t)=\alpha_0 + 
\alpha^\prime t$ leads to a \wgp\ dependence \siggp~$\propto 
\wgpw^{4\,(\alpha_{0}-1)}/b(\alpha^\prime,\ln\wgpw)$ \cite{collins,1pom}.
The parameters of the established soft 
and the proposed hard pomeron trajectories are 
$(\alpha_0,\alprim)_{\mbox{soft}}=(1.08,0.25\,\gevmsq)$ and 
$(\alpha_0,\alprim)_{\mbox{hard}}=(1.418,0.1\,\gevmsq)$~\cite{2pom}. By using these 
trajectories and fitting the relative contributions and mixing as defined in 
\cite{2pom}, a good representation of the data is given. The 
relative contributions of hard, soft and mixing terms are found to vary between 0.1 : 
0.5 : 0.4 at \wgp$=30\,\gev$ and 0.5 : 0.1 : 0.4 at \wgp$=250\,\gev$.

%%%%%%%%%%%%%%%%%%%%%%%%%%%%%%%%%%%%%%%%%%%%%%%%%%%%%%%%%%%%%%%%%%%%%%%%%%%%%%%%%%%%%%%%%%%%%%%%%%%

\section{Differential Cross Section \protect\boldmath\dsdt\ for \boldmath{$J/\psi$} 
Mesons}

The dependence of the photoproduction cross section of  \jpsi\ mesons on $t$ is 
studied in the region of  $40\leq\wgpw\le 150\,\gev$, which provides good $t$ 
resolution. First, this dependence is measured averaging over the entire 
$W_{\gamma p}$ range, assuming an exponential behaviour. Then, the dependence of 
\dsdt\ on \wgp\ in fixed $t$ bins is determined and an effective Regge trajectory is 
derived.

%%%%%%%%%%%%%%%%%%%%%%%%%%%%%%%%%%%%%%%%%%%%%%%%%%%%%%%%%%%%%%%%%%%%%%%%%%%%%%%%%%%%%%%%%%%%%%%%%%%

\paragraph*{\boldmath $t$ Distribution:} 

\begin{figure}[t!] \centering
  \setlength{\unitlength}{1cm}
  \begin{picture}(16.,8.)
    \put(1.,-1.2){\epsfig{file=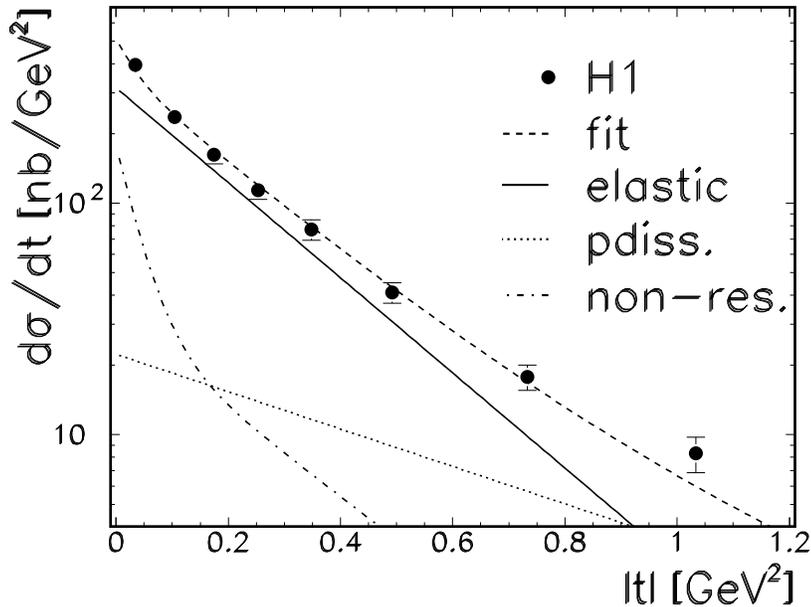,width=13cm}}
  \end{picture}
\caption{\em The differential cross section $d\sigma/dt$ for elastic $J/\psi$ 
         photoproduction, as a function of $|t|$, averaged over the energy region 
         $40\leq W_{\gamma p}\leq 150\:\mbox{GeV}$. Only statistical errors are given. 
         Also shown is the result of a fit (dashed line; $\chi^2/\mathit{dof}=4.7/6$) 
         to the data which includes the three different contributions described in the 
         text. The solid curve shows the fit result for elastic \protect\jpsi\ 
         production assuming an exponential $t$ dependence.}
  \label{fig3}
\end{figure}

The variable $t$ is approximated by $t \simeq -(\vec{p}_{t})^2$, where $\vec{p}_{t}$ 
is the transverse momentum of the $J/\psi$ meson.
%The resolution, obtained from the 
%Monte Carlo simulation for the reconstruction of $t$, ranges from about $25\%$ at 
%$|t|<0.1\,\gevt$ to $14\%$ at $0.5<|t|<1.2\,\gevt$.
The resolution of $t$, obtained from the Monte Carlo simulation of 
vector mesons, is typically 
$0.035\,\gevt$ and increases to $0.060\,\gevt$ for  $|t|\approx 1.2\,\gevt$.
The events without a signal in 
the forward detectors, in the mass range $2.9\leq M_{\mu\mu}\leq 3.3\,\mbox{GeV}$, 
are divided into bins in $t$ and corrected for efficiencies but not for remaining 
backgrounds. Photon-proton cross sections are calculated as described in 
section~\ref{crosssection}, yielding the resulting values of \dsdt\ shown in 
Figure~\ref{fig3}. 

With this selection, the data contain elastic \jpsi\ events ($81\%$), events with proton 
dissociation ($13\%$) and non-resonant background ($6\%$). These fractions result from  
the analysis described in section~\ref{crosssection} and are fixed in the subsequent 
analysis of the $t$-dependence. The $t$-dependence of the proton dissociative \jpsi\ 
contribution is obtained by fitting those corrected data which contain a signal in 
the forward detectors with one exponential function. The different $t$ dependence of the  
efficiencies for proton dissociative events with and without signals in the forward 
detectors is taken into account. The non-resonant background is described by a sum of 
two exponential functions which are obtained by fitting the data outside of the \jpsi\ 
mass region. 

A common fit of the three contributions to the data is then carried out, assuming an 
exponential distribution $e^{-b\,|t|}$ for the elastic \cs\ (see Figure~\ref{fig3}). 
The only free parameters in this fit are the elastic slope parameter $b$ and an overall 
normalization. Only the statistical errors of the data are taken into account in the fit. 
A value $b=(4.73\pm0.25^{+0.30}_{-0.39})\,\gmt$ is derived for $|t|\leq 1.2\,\gevt$, where 
the first error is statistical. The systematic uncertainty is obtained by varying the data 
selection cuts, the $t$ distribution of the generated events for both the acceptance and 
efficiency corrections and the admixture of the background contributions (both relative 
normalizations and shapes). The resulting value of $b$ agrees within errors with the 
previous H1 result~\cite{H1962} and with a similar result from the ZEUS 
collaboration~\cite{Zeus971}. 

%%%%%%%%%%%%%%%%%%%%%%%%%%%%%%%%%%%%%%%%%%%%%%%%%%%%%%%%%%%%%%%%%%%%%%%%%%%%%%%%%%%%%%%%%%%%%%%%%%%

\paragraph*{Regge Trajectory:}

In the description of elastic scattering based on Regge phenomenology and pomeron 
exchange, the energy dependence of the elastic \cs\ follows a power law:
\begin{equation}
\frac{d\sigma}{d t}=\frac{d\sigma}{d t}\bigg|_
              {\substack{t=0,}{\wgpw=W_0}}\cdot e^{-b_0\, |t|}
           \bigg(\frac{W_{\gamma p}}{W_0}\bigg) ^{4(\alpha(t)-1)}, \label{eqtr}
\end{equation}
where $\alpha(t)= \alpha_0 + \alpha^\prime t$ denotes the exchanged trajectory and $b_0$ 
and $W_0$ are constant parameters not determined by the theory. Whereas Regge theory 
predicts a change of the \wgp\ dependence with $t$, there is no such simple relationship 
in pQCD~\cite{F_K_S}.

\begin{figure}[b!] \centering
  \setlength{\unitlength}{1cm}
  \begin{picture}(16.,10.5)
    \put(-0.7,-1.0){\epsfig{file=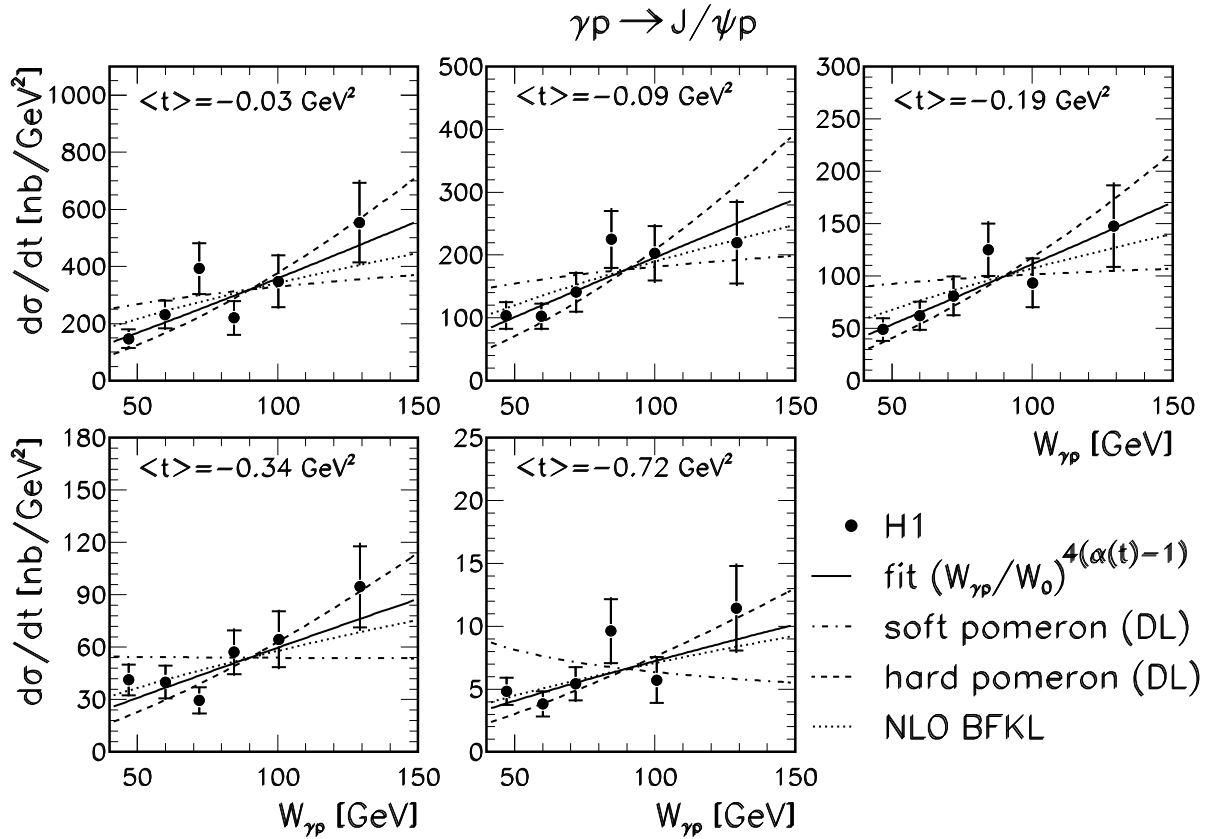,width=19cm}}
  \end{picture}
  \caption{\em The differential cross section $d\sigma/dt$ as a function of 
           $W_{\gamma p}$ in five bins of $t$ together with a fit of the form 
           $d\sigma/dt=N\cdot(W_{\gamma p}/W_0)^{4(\alpha(t)-1)}$ (solid line). 
           The inner error bars on the data points show the dominating statistical 
           errors and the outer bars the total error. The predictions of the soft 
           and hard Donnachie-Landshoff pomeron trajectories~\protect\cite{2pom} 
           are shown as dash-dotted 
           and dashed lines. The prediction based on a NLO BFKL 
           calculation~\protect\cite{bfklpom} is given 
           as a dotted line. All theoretical curves are normalized to the fit at 
           $W_0=90\,$GeV.} 
  \label{fig4}
\end{figure}

In order to determine the effective trajectory $\alpha(t)$ in elastic $J/\psi$ 
production, the data used for Figure~\ref{fig3} are analysed as functions of \wgp\ 
and $t$. The data are divided into five bins in $t$ and six bins in \wgp. According to 
the Monte Carlo simulation, the migrations due to resolution effects range from 
$20-50\%$ in $t$ and are negligible in \wgp. The differential \cs\ $d\sigma/dt$\ is 
determined in each bin ($t_i,W_j$), as described in 
section~\ref{crosssection}, i.e. correcting for non resonant and 
proton dissociative background. The 
resulting values are listed in Table~\ref{tab3} and shown in Figure~\ref{fig4} 
as functions of \wgp\ in the five $t$ bins. Statistical and systematic errors, excluding 
the contributions which affect only the normalization, are taken into account; the 
statistical errors dominate. The data in each bin  $t_i$ are fitted as a function of 
\wgp\ (Equation~\ref{eqtr}) with $\alpha(t_i)$ and a normalization constant as free 
parameters.

The results of these fits are shown as full lines in Figure~\ref{fig4}. The respective 
curves, corresponding to the soft and the hard pomeron trajectories by Donnachie and 
Lands\-hoff~\cite{2pom} as well as a hard pomeron extracted from a NLO BFKL 
calculation~\cite{bfklpom}, are also shown. Note that the differential \cs\ \dsdt\ at 
fixed $t$ exhibits a rise with \wgp\ at all $t$ values, in contrast to the expectation 
from the soft pomeron model.

\begin{figure}[t!] \centering
  \setlength{\unitlength}{1cm}
  \begin{picture}(16.,11.)
    \put(2.5,-1.7){\epsfig{file=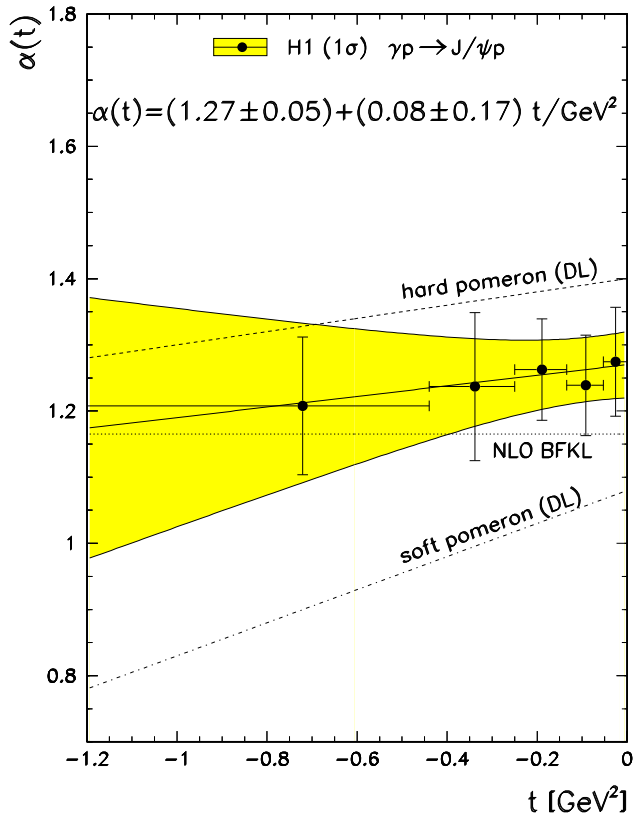,width=10cm}}
  \end{picture}
  \caption{\em The measured Regge trajectory \protect{$\alpha(t)=\alpha_0+\alprim t$} 
           for the process $\gamma p \rightarrow J/\psi\,p$. The solid line shows the 
           result of the fit. The one standard deviation contour is indicated by a 
           shaded band. Also shown are the soft and the hard Donnachie-Landshoff 
           pomeron trajectories~\protect\cite{2pom} and a result based on a NLO BFKL 
           calculation~\protect\cite{bfklpom}.}
  \label{fig5}
\end{figure}

The resulting fit values of $\alpha(t_i)$ are shown as points in Figure~\ref{fig5}; 
the error bars contain the statistical and systematic uncertainties. A fit to these 
values of the form $\alpha(t)= \alpha_0 + \alpha^\prime t$, where $\alpha_0$ and 
$\alprim$\ are free parameters, yields the result
$$ \alpha(t)=(1.27\pm0.05)+(0.08\pm0.17)\cdot t/\gevsq .$$

This fit is shown in Figure~\ref{fig5}, including a band which reflects the one standard 
deviation uncertainty, taking into account correlations between $\alpha_0$ and 
$\alpha^\prime$. The resulting effective trajectory lies, as expected from the analysis 
of the total \cs\ (see Figure~\ref{fig2}a), in between the soft and the hard pomeron 
trajectories of Donnachie and Landshoff. A prediction for a pomeron, derived from a 
NLO BFKL calculation~\cite{bfklpom}, is also shown in Figure~\ref{fig5}. Note, however, 
that in this calculation only $\alpha_0$ is predicted\footnote{We chose the solution 
corresponding to $\xi=3$ in Table 2 of \cite{bfklpom}.} and $\alpha^\prime=0$ is assumed 
for the plot. The measured intercept $\alpha_0$ in $J/\psi$ production is found to be 
significantly larger than in the production of the light vector mesons $\rho$ and $\phi$. 
The slope $\alpha^\prime$ is found to be compatible with zero, in contrast to the value 
measured for $\rho$ and $\phi$ mesons~\cite{Zeus992}. The present results allow, for the 
first time, the extraction of the effective Regge trajectory exchanged in $J/\psi$ 
photoproduction from a single experiment, avoiding relative normalization uncertainties 
between experiments. A previous fit~\cite{levy}, using earlier HERA results and fixed 
target data at lower \wgp, gives compatible results.

%%%%%%%%%%%%%%%%%%%%%%%%%%%%%%%%%%%%%%%%%%%%%%%%%%%%%%%%%%%%%%%%%%%%%%%%%%%%%%%%%%%%%%%%%%%%%%%%%%%

\section{\boldmath{\ups} Production}

\begin{figure}[b!] \centering
   \setlength{\unitlength}{1cm}
   \begin{picture}(17.,6.3)
     \put(-1.5,-2.){\epsfig{file=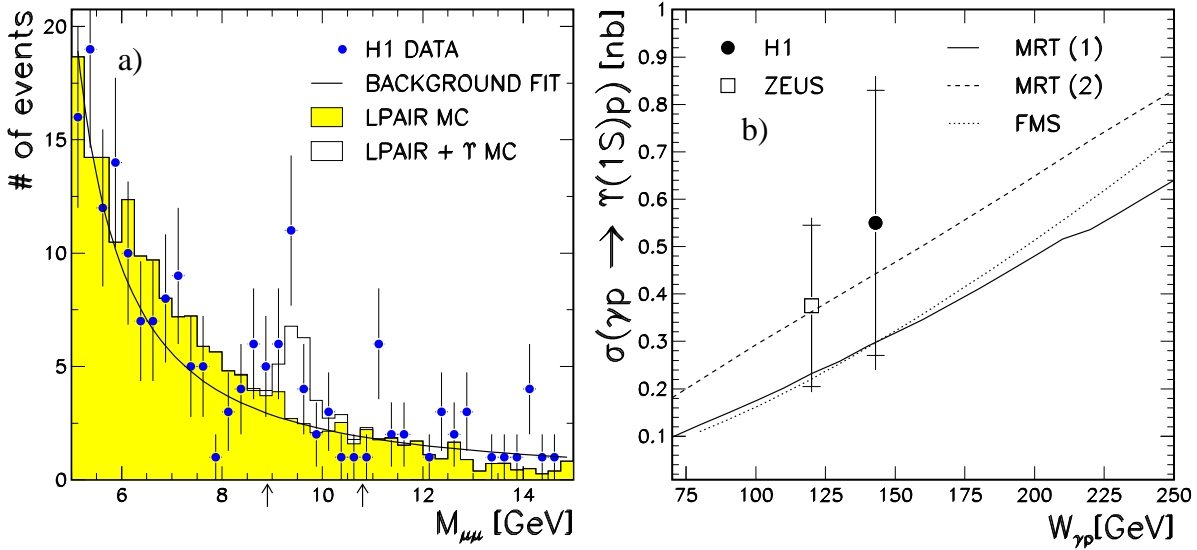,width=18.5cm}}   
   \end{picture}
 \caption{\em a) The invariant mass spectrum of the selected di-muon events 
    without a signal in the forward detectors. The solid line shows the 
    result of a fit to the background region. The histograms are Monte Carlo 
    predictions for \protect\ggmm (LPAIR) and for the three lowest $\Upsilon$ 
    states (DIFFVM, normalized to the observed number of events). The arrows 
    indicate the chosen signal region. 
    b) The cross section for elastic photoproduction of 
    $\Upsilon(1S)$ from this analysis with the ZEUS result (which also assumes a 
    $70\%$ contribution of the $1S$ state). The inner and outer error bars show 
    the statistical and the total errors, respectively. Predictions based on pQCD, 
    calculated with the gluon distribution MRSR2~\protect\cite{MRSR}, are shown. 
    MRT(1)~\protect\cite{MRT99} and FMS~\protect\cite{F_S_M99} are calculations 
    based on the leading order vector meson cross section, including corrections. 
    MRT(2)~\protect\cite{MRT99} employs parton hadron duality to derive the 
    prediction for the $\Upsilon$ from the $b\bar{b}$ cross sections.}
  \label{fig6}
 \end{figure}

The mass distribution of the selected di-muon events, without a signal in the forward 
counters, is shown in Figure~\ref{fig6}a. An excess of events is visible at a 
mass of $\sim9.4\,\gev$ above a background which is fitted by a power law. The background 
is, within errors, described by the process \ggmm. The significance of the excess is 
about two standard deviations and its position is compatible with these events arising 
from \upsos\ decay. The mass resolution is $\sigma\sim 250\,$MeV. Due to limited 
statistics, small contributions from the two lowest radial excitations \upsts\ and 
\upsths\ cannot be excluded. In order to estimate the number of signal events, a wide 
mass interval is chosen which extends from $2\,\sigma$ below the \upsos\ to $2\,\sigma$ 
above the \upsths\ state ($8.9\lsim m_{\mu\mu}\lsim 10.8\:$GeV). Subtracting the 
background, $12.2\pm 6.3$ signal events are obtained. Corrections for the remaining 
background, due to diffractive proton dissociation ($16\%$) and the total selection and 
trigger efficiency ($24\%$), are determined from the Monte Carlo simulation, cross checked 
with data similarly to the $J/\psi$ analysis. Taking into account the total integrated 
luminosity and using the \wwa~\cite{WWA}, a \gpcs\ in the range $70\leq\wgpw\leq250\,\gev$ 
with an average value of $\langle\wgpw\rangle=143\,\gev$ is obtained:
$$\sigma(\gamma\:p\rightarrow \Upsilon\:p)\times
         \mbox{BR}(\Upsilon\rightarrow\mu^+\mu^-)=(19.2\pm9.9\pm4.8)\:\mbox{pb},$$
 
where \ups\ includes the states \upsos, \upsts\ and \upsths. The first error is statistical 
and the second systematic ($25\%$). The latter is calculated by varying the definition of 
the signal and background regions, using different estimates of the background function 
or calculating the background from the \ggmm\ prediction and taking into account 
uncertainties of efficiencies.

In order to extract the \cs\ for the production of the \upsos\ state, an assumption must 
be made about its relative production ratio. We choose a fraction of $70\%$. This is in broad 
agreement with an estimate using the branching ratios and the electronic widths of the 
different $\Upsilon$ states~\cite{PDG98} and with a recent calculation~\cite{F_S_M99}. 
Using the standard branching ratios for \upsos, \upsts\ and \upsths~\cite{PDG98}, 
the result for the elastic $\Upsilon(1S)$ photoproduction cross section is 
\begin{equation*}
\sigma(\gamma p\rightarrow\Upsilon(1S) p) = (0.55\pm0.28\pm0.14)\:\mbox{nb}\:\mbox{.}
\end{equation*}

In Figure~\ref{fig6} b), the measured cross section $\sigma(\gamma 
p\rightarrow\Upsilon(1S) p)$ is shown together with the ZEUS measurement~\cite{ups_zeus}, 
which agrees well with our result.  Recent pQCD calculations~\cite{F_S_M99,MRT99} are 
able to describe the data after consideration of two effects: the real part of the 
scattering amplitude and the non-diagonal parton distributions in the proton. Such 
effects are found to be more important for the production of the \ups\ than for the 
$J/\psi$ meson, due to its larger mass.

%%%%%%%%%%%%%%%%%%%%%%%%%%%%%%%%%%%%%%%%%%%%%%%%%%%%%%%%%%%%%%%%%%%%%%%%%%%%%%%%%%%%%%%%%%%%%%%%%%%

\section{Summary}

Improved results over an extended kinematic range on the elastic photoproduction of 
\jpsi\ mesons are presented. The total photoproduction cross section is measured as a 
function of the photon-proton \cms\ energy \wgp\ in the range $26\leq\wgpw\leq 285\,\gev$, 
where it can be para\-meterized as $\sigma_{\gamma p} \propto W_{\gamma p}^\delta$ with 
$\delta=0.83\pm0.07$. 

The data are well described by a perturbative QCD calculation by Frankfurt et
al.~\cite{F_K_S97}, where a good description is obtained using the CTEQ4M or MRSR2 gluon 
densities in the proton. The data in this \wgp\ range can also be described by a 
non-perturbative model which assumes the exchange of two pomeron trajectories. However, a 
model with only a soft pomeron is ruled out. 

The differential \cs, \dsdt, for elastic \jpsi\ production, averaged over the range 
$40\leq W_{\gamma p}\leq 150\;\mbox{GeV}$, is well described by a single exponential 
for the accessible range below $|t|=1.2\gevt$ with a slope parameter 
$b=(4.73\pm0.25^{+0.30}_{-0.39})\,\gmt$.

In the same \wgp\ range, $d\sigma/dt$ is measured as a function of \wgp\ and $t$. 
From this analysis, the parameters of the effective trajectory for elastic $J/\psi$ 
photoproduction are determined using data from this experiment only. The values of the 
intercept and the slope of the measured trajectory lie between those of the well known 
soft and the conjectured hard pomeron trajectories of Donnachie and Landshoff. The 
measured intercept is compatible with a prediction derived from a NLO BFKL calculation. 

Elastic photoproduction of \ups\ mesons is observed. The \cs\ for the sum of the three 
lowest states is measured. The elastic photoproduction \cs\ for \upsos\ mesons is 
extracted. Recent calculations within pQCD are in agreement with this value. 

%%%%%%%%%%%%%%%%%%%%%%%%%%%%%%%%%%%%%%%%%%%%%%%%%%%%%%%%%%%%%%%%%%%%%%%%%%%%%%%%%%%%%%%%%%%%%%%%%%%

\section*{Acknowledgements}

We are grateful to the HERA machine group, whose outstanding efforts made this experiment 
possible. We appreciate the immense effort of the engineers and the technicians who 
constructed and maintained the detector. We thank the funding agencies for their financial 
support of the experiment. We wish to thank the DESY directorate for the support and 
hospitality extended to the non-DESY members of the collaboration. We also thank 
M.\,McDermott and T.\,Teubner for useful discussions and for providing us with their 
model predictions, W.\,Koepf for making his program available and H.\,Spiesberger for 
discussions on radiative corrections.

%%%%%%%%%%%%%%%%%%%%%%%%%%%%%%%%%%%%%%%%%%%%%%%%%%%%%%%%%%%%%%%%%%%%%%%%%%%%%%%%%%%%%%%%%%%%%%%%%%%
\begin{table}[p]\centering
{\normalsize 
\begin{tabular}{|rcl|c||rll|} \hline
  \multicolumn{3}{|c|}{\wgp\ ~interval $[\mbox{GeV}]$} & $\langle W_{\gamma p}\rangle\ [\mbox{GeV}]$\ \ &
      \multicolumn{3}{|c|}{$\sigma (\gamma p \rightarrow J/\psi p)\ [\mbox{nb}]$} \\ \hline
  
   26 & -- & 36\ \   & 31.0  \  & \ \ 19.8  & $\pm$ 3.7  & $\pm$ 5.1 \\
   40 & -- & 52\ \   & 46.2  \  & \ \ 29.8  & $\pm$ 3.7  & $\pm$ 3.4  \\
   52 & -- & 60\ \   & 56.1  \  &     41.7  & $\pm$ 5.2  & $\pm$ 4.8  \\
   60 & -- & 67.8\ \ & 64.0  \  &     49.5  & $\pm$ 6.6  & $\pm$ 5.6  \\
 67.8 & -- & 77\ \   & 72.5  \  &     51.8  & $\pm$ 6.9  & $\pm$ 5.9  \\
   77 & -- & 86\ \   & 81.5  \  &     62.4  & $\pm$ 8.5  & $\pm$ 7.1  \\
   86 & -- & 97\ \   & 92.0  \  &     67.6  & $\pm$ 8.8  & $\pm$ 7.7  \\
   97 & -- & 113\ \  & 105.2  \ &     64.6  & $\pm$ 9.6  & $\pm$ 7.4  \\
  113 & -- & 150\ \  & 133.4  \ &     89.0  & $\pm$ 13.2 & $\pm$ 12.7 \\
  135 & -- & 160\ \  & 147.3  \ &     80.6  & $\pm$  7.9 & $\pm$ 12.7 \\ 
  160 & -- & 185\ \  & 172.2  \ &    104.0  & $\pm$  7.9 & $\pm$ 16.4 \\
  185 & -- & 210\ \  & 197.1  \ &    113.6  & $\pm$ 12.2 & $\pm$ 18.0 \\ 
  210 & -- & 235\ \  & 222.4  \ &    116.0  & $\pm$ 14.8 & $\pm$ 18.5 \\ 
  235 & -- & 260\ \  & 247.4  \ &    118.9  & $\pm$ 15.6 & $\pm$ 19.0 \\
  260 & -- & 285\ \  & 272.4  \ &    156.9  & $\pm$ 20.3 & $\pm$ 25.0 \\ \hline

\end{tabular}
\caption{\em Cross sections for the elastic process $\gamma p \rightarrow J/\psi\: p$ in bins of
  \wgp. The first error of the cross section is the statistical error and the second is the 
systematic uncertainty.} 
\label{tab2}}
\end{table}

\begin{table}\centering
{\normalsize 
\renewcommand{\arraystretch}{1.1} 
$\begin{array}{|c||c|c|c|c|c|c|}\hline
   & \multicolumn{6}{|c|}{\wgpw\ \mbox{interval}~[\mbox{GeV}]} \\  
-t~[\gevsq]      &  40-54     &  54-66     &  66-78     &  78-91     &  91-110    &  110-150    \\  \hline\hline
0-0.053        &  146.7     &  231.6     &  392.4     &  220.1     &  348.2     &  553.5      \\ 
(0.03)         &  \pm 32.5  &  \pm 48.4  &  \pm 89.3  &  \pm 58.7  &  \pm 90.1  &  \pm 138.7  \\
                &  \pm 13.4  &  \pm 21.1  &  \pm 35.7  &  \pm 20.0  &  \pm 31.7  &  \pm 58.1   \\ \hline
0.053-0.134    &  103.8     &  102.6     &  140.7     &  224.9     &  202.6     &  219.4      \\
(0.09)         &  \pm 21.3  &  \pm 20.3  &  \pm 80.9  &  \pm 45.2  &  \pm 43.3  &  \pm 64.9   \\
                &  \pm 9.4   &  \pm 9.3   &  \pm 12.8  &  \pm 20.5  &  \pm 18.4  &  \pm 23.0   \\ \hline
0.134-0.25     &  48.9      &  62.0      &  80.9      &  125.2     &  93.4      &  147.6      \\
(0.19)         &  \pm 10.9  &  \pm 13.6  &  \pm 18.6  &  \pm 25.0  &  \pm 23.3  &  \pm 38.9   \\
                &  \pm 4.4   &  \pm 5.6   &  \pm 7.4   &  \pm 11.4  &  \pm 8.5   &  \pm 15.5   \\ \hline
0.25-0.44      &  41.3      &  40.0      &  29.5      &  57.2      &  64.6      &  94.6       \\
(0.34)         &  \pm 8.8   &  \pm 9.4   &  \pm 7.5   &  \pm 12.6  &  \pm 16.0  &  \pm 23.1   \\
                &  \pm 3.8   &  \pm 3.6   &  \pm 2.7   &  \pm 5.2   &  \pm 5.9   &  \pm 9.9    \\ \hline
0.44-1.2       &  4.82      &  3.82      &  5.42      &  9.62      &  5.72      &  11.43      \\ 
(0.72)         &  \pm 1.09  &  \pm 0.99  &  \pm 1.33  &  \pm 2.54  &  \pm 1.81  &  \pm 3.35   \\
                &  \pm 0.44  &  \pm 0.35  &  \pm 0.49  &  \pm 0.88  &  \pm 0.52  &  \pm 1.20   \\ \hline
\end{array}$
\setlength{\unitlength}{0.5cm}

\caption{\em Differential cross sections $d\sigma/dt\:$[nb/GeV$^2$] for the 
         elastic process 
         $\gamma p \rightarrow J/\psi\: p$ in bins of $t$ and \wgp. 
         The $t$ intervals and bin centres are given in the left column. The first error 
          is 
         statistical and the second is the systematic uncertainty neglecting pure 
         normalization errors.} 
\label{tab3}}
\end{table}

%%%%%%%%%%%%%%%%%%%%%%%%%%%%%%%%%%%%%%%%%%%%%%%%%%%%%%%%%%%%%%%%%%%%%%%%%%%%%%%%%%%%%%%%%%%%%%%%%%%

%%%%%%%%%%%%%%%%%%%%%%%%%%%%%%%%%%%%%%%%%%%%%%%%%%%%%%%%%%%%%%%%%%%%%%%%%%%%%%%%%%%%%%%%%%%%%%%%%%%

\end{document}